\documentclass[12pt,letterpaper,oneside,onecolumn]{article}
\usepackage[]{fontenc}
\usepackage[dvips]{epsfig}

\makeatletter
\def\LyX{L\kern-.1667em\lower.25em\hbox{Y}\kern-.125emX\spacefactor1000}%
\newcommand{\lyxtitle}[1] {\thispagestyle{empty}
\global\@topnum\z@
\section*{\LARGE \centering \sffamily \bfseries \protect#1 }
}

\newcommand{\lyxletterstyle}{
\setlength\parskip{0.7em}
\setlength\parindent{0pt}
}

\makeatother

\lyxletterstyle
\begin{document}

{\centering \bfseries \large Determination of \em When \em an
Outcome is Actualised in a Quantum Measurement using DNA-Photolyase
System \par}

\( \smallskip  \)

\( Dipankar \) \( Home^{*},Rajagopal \) \( Chattopadhayaya^{**} \)

*Department of Physics, Bose Institute, Calcutta 700009, India; dhom@boseinst.ernet.in

**Department of Biochemistry, Bose Institute, Calcutta 700054, raja@bic.boseinst.ernet.in

\bfseries The biochemical attachment of photolyase to ultraviolet
(uv) absorbed DNA molecules provides a method for registering whether
a source has emitted photons. Here using laws of chemical kinetics
and related experimental methods we argue that \em \em the \em instant
\em after which this information becomes \em discernible \em can be
\em empirically determined \em by retrodicting from relevant data
\em when \em the photolyase binding to uv-absorbed DNA molecules has
started occurring. Thus an empirically investigable twist is provided
to the quantum measurement problem. \mdseries 

\( \smallskip  \)

{\centering --------\par}

If a system is initially in a state \( \psi (\psi =a\psi _{1}+b\psi _{2}) \) which is a superposition of
two states \( \psi _{1} \)and \( \psi _{2} \) that are eigenstates of a dynamical variable which
is measured, a general characteristic of its interaction with a measuring
device is that it results in a final state of the form 

\hfill{}\( \Psi =a\psi _{1}\Phi _{1}+b\psi _{2}\Phi _{2} \) \hfill{}(1) 

where \( \Phi _{1} \)and \( \Phi _{2} \) are mutually orthogonal and \em macroscopically distinguishable
\em states of the device. It is an ineluctable feature of linear unitary
quantum mechanical treatment of \em any \em measurement process that
the final state of system coupled to measuring apparatus has an entangled
nonfactorisable form given by Eq.(1). 

The much debated quantum measurement problem [1-8] stems from the
\em meaning \em of a \em pure state \em wave function in quantum mechanics
giving rise to an inherent \em incompatibility \em between a wave
function of the form (1) and \em actualisation \em of the result of
a measurement. A pure state in quantum mechanics means that each member
(in this case, a system coupled to an apparatus) of an ensemble described
by a pure state \( \Psi  \) as given by Eq. (1) has the \em same \em wave function
\( \Psi  \). Thus a pure state in quantum mechanics corresponds to a \em homogeneous
\em ensemble whose members are \em indistinguishable\em . On the other
hand, \em definiteness \em of individual outcomes requires that different
outcomes are \em distinguishable\em . All measurements culminate in
the final ensemble of systems coupled to apparatus which is essentially
\em heterogeneous\em . A heterogeneous ensemble is, however, represented
by a \em mixed state \em in quantum mechanics. Since within standard
quantum mechanics under no unitary time evolution a pure state can
evolve into a mixed state (see, for instance, [7] pp. 87-88), \em how
\em to coherently accommodate within quantum mechanics the occurrence
of distinguishable outcomes is thus an intriguing ``paradox''. Not surprisingly,
Weinberg [9] has called this ``the most important puzzle in the interpretation
of quantum mechanics.'' 

Resolution of the measurement problem requires either modifying or
enlarging the standard framework of quantum mechanics in an appropriate
way [10]. Schemes to this end are of two categories : (a) Approaches
[11-15] which leave the mathematical formalism of quantum mechanics
unmodified but introduce new elements into the conceptual framework.
(b) Models [16-19] which modify the mathematical formalism (preserving
the usual results of the standard formulation in their empirically
verified domains) in order to provide a dynamical description of a
measurement process in terms of an actual transition from a pure to
a mixed state (the so-called ``collapse of a wave function''). 

A key ingredient in any of the above approaches is the notion that
a measurement outcome has a definite \em objective \em (observer-independent)
\em reality \em in the sense that once \em actualised\em , it remains
``out there'' so that the result can be inspected at \em any \em subsequent
instant without perturbing it. This therefore requires a precise specification
of \em at what stag\em e a measurement result is recorded in a \em stable
\em and \em discernible \em form. The schemes mentioned above use
different \em a priori \em criteria for specifying this stage. An
objective formulation of such a criterion amenable to empirical scrutiny
is \em necessary \em for developing a satisfactory resolution of the
measurement problem. On this question the present paper indicates
a new direction of study by using biomolecular analogues of quantum
measuring devices which promise to be of particular significance because
biomolecules such as DNA ``occupy a strategic position between microscopic
and macroscopic bodies''[20]. 

The specific example considered in this paper pertains to ultraviolet
(uv) absorption by DNA molecules which develop new covalent bonds
at certain sites (formation of uv-induced pyrimidine dimers) leading
to a global readjustment of atomic positions. The displacement of
an individual atom relative to its neighbouring atoms at such a damaged
site is \( \leq 2\times 10^{-8}cm \) which cannot \em \em be directly observed by using an electron
microscope. However, it becomes discernible by means of a curious
biochemical property, viz. that if there are nearby \em photolyase
enzyme \em molecules, some of them get preferentially attached to
the damaged sites in the uv-absorbed DNA (uv-induced pyrimidine dimers
in DNA have a chemical affinity towards photolyase which is a single
polypeptide chain of 454 to 614 amino acids). That such an arrangement
registering arrival of uv photons provides an instructive example
of quantum measurement was earlier discussed by us [21]. We pointed
out specific characteristics of this example in the context of the
measurement problem. The central point is that since uv-damaged DNA
molecule attached to photolyase records information about the emission
of photons in a form that can be inspected at will any number of times
at arbitrary instants, it is not essentially different from any readout
device registering a measurement outcome. 

The present paper lends a new twist to the above example by invoking
suitably chosen features of \em chemical kinetics \em associated with
the biochemical process of photolyase enzyme binding to uv-absorbed
DNA molecules. We argue that it is possible to estimate retrodictively
from suitable measurements \em when \em the enzyme attachment process
began (say, at \( t_{0} \)). Note that \( t_{0} \) signifies the instant from which information
about the relevant measurement outcome (viz. that a source has emitted
photons) is available in a \em stable \em and \em discernible \em form.
That is, if one chooses to make a measurement at \em any \em \( t\geq t_{0} \), some
DNA molecules can always be found attached to photolyase thereby indicating
that uv photons had been absorbed. However, as already mentioned,
\em no \em physical significance can be associated with \( t_{0} \) within standard
quantum mechanics in the sense of signifying the onset of an actual
transition from a pure to a mixed state. This example therefore helps
to sharpen the quantum measurement problem by bringing it out in an
empirically relevant form. 

The measurement under consideration is as follows. Given a source
which has a probability of emitting a pulse of, say, \( 10^{9}\gamma  \) photons (as
explained later, the reason for choosing this size of the pulse is
for convenience in outlining a concrete feasible experimental arrangement
), the measurement in question is designed to find out not only \em whether
\em the source (left to itself for a certain time) has emitted such
a pulse but also that \em if \em a pulse is emitted, one can find
out from \em which \em instant onwards information about the pulse
emission becomes available. In the first stage, the emitted pulse
of \( 10^{9}\gamma  \) photons interact with a pure CsI crystal proliferating into a
pulse of \( 10^{15} \)uv photons . This output pulse is still \em transient \em which
in itself does \em not \em constitute a stable record of measurement
information. To convert this into a \em stable \em record, these uv
photons need to be incident on a ``detecting device'' which in our case
is an aqueous solution of DNA mixed with photolyase. 

The combined state of source coupled with the ``detecting device'' mentioned
above is given by ( a particular form of Eq.(1) ) 

\hfill{}\( \Psi =a\psi _{e}\Phi _{A}+b\psi _{0}\Phi _{0} \)\hfill{} (2)

where \( \psi _{e} \), \( \psi _{0} \) denote states of the source corresponding to emission and
no emission and \( \Phi _{A} \), \( \Phi _{0} \) denote photolyase attached and unattached states
of DNA respectively. Note that \( \left| a\right| ^{2} \), \( \left| b\right| ^{2} \) denote respectively the probabilities
of the source emitting or not emitting photons within the specified
time interval. While writing Eq.(2) we assume that the aqueous solution
in this detecting arrangement comprises sufficient number of DNA and
photolyase molecules so that in the event of the source emitting photons,
an appreciable number of DNA molecules absorbing the photons get attached
to photolyase. This binding occurs through a \em time-evolving \em chemical
process. Here the role of photolyase is crucial for forming \em macroscopically
distinguishable \em states \( \Phi _{A},\Phi _{0} \). There are of course other methods as
well for distinguishing between uv damaged and undamaged DNA (e.g.,
using nuclear magnetic resonance, x-ray crystallography or relevant
chemical properties of other enzymes), but for our purpose the use
of photolyase appears most convenient. 

The crucial point is that (as we have argued earlier) Eq. (1) or Eq.
(2) does \em not \em in itself account for the registration of a definite
outcome. If one accepts the ``completeness'' of a wave function in specifying
the state of an individual system, an \em additional \em \em hypothesis
\em of the collapse of a wave function from a pure to a mixed state
is required in order to explain the emergence of a measurement outcome.
At which precise stage this putative transition occurs is an inherently
contentious issue. In our specific example two different viewpoints
[21,22] are possible :

(a) The actual collapse of a pure state wave function (Eq.(2)) to
a mixture of states \( \psi _{e}\Phi _{A},\psi _{0}\Phi _{0} \) occurs \em only when \em an observation is made
by an external means to find out whether some DNA molecules have got
attached to photolyase. 

(b) Collapse of the wave function (Eq.(2)) into a mixture \( \psi _{e}\Phi _{A},\psi _{0}\Phi _{0} \) begins
to take place at the instant \( t_{0} \)when one or more DNA molecules in the
solution \em start \em getting attached to photolyase, independent
of whether being observed by external means. This point of view is
motivated on the ground that at \( t_{0} \) an initially homogeneous ensemble
actually begins to split into two physically distinct subensembles
comprising photolyase bound and unbound DNA respectively. Hence \( t_{0} \)
has an objective (observer-independent) significance that needs to
be incorporated within the description of the collapse of a wave function.

We shall now indicate how it is possible to estimate from appropriately
collected empirical data the instant \( t_{0} \) photolyase binding to DNA begins.
In other words, the \em instant \em from which information about the
photon emission can be known (discernible by any external observer)
is shown to be \em empirically determinable\em . Thus, in this specific
context, the viewpoint (a) appears to be untenable. How to accommodate
the viewpoint (b) within the various suggested forms of dynamical
models of wave function collapse is a nontrivial issue [21,22] which
needs to be studied in detail. Even for models addressing the measurement
problem without using the idea of wave function collapse, this type
of study involving biomolecular systems as measuring devices should
provide useful constraints about specifying the stage at which a measurement
outcome is recorded. 

We now proceed to discuss specifics of the relevant experimental scheme.
The rate of formation of uv-damaged \em DNA-photolyase complex \em (whose
instantaneous concentration is denoted by \( [PS]_{t} \)) depends on the instantaneous
concentrations of potential enzyme attachment sites in uv-damaged
DNA denoted by \( [S]_{t} \) as well as that of free or unbound photolyase denoted
by \( [P]_{t} \). This biochemical reaction follows the second order rate law

\hfill{}\( d[PS]_{t}/dt=k[P]_{t}[S]_{t} \) \hfill{}(3) 

where \( [P]_{t}=P_{0}-[PS]_{t} \) with \( P_{0} \) being the initial concentration of photolyase, \( [S]_{t}=S_{0}-[PS]_{t} \) where
\( S_{0} \) is the initial concentration of potential enzyme attachment sites
produced in uv-damaged DNA and k is the second order rate constant.
Phenomenological basis of Eq.(3) lies in random collision between
uv-damaged DNA and photolyase arising from a diffusion process in
the aqueous solution which ultimately result in specific chemical
binding between the two. It is \em experimentally verified \em [23]
that the uv-damaged DNA-photolyase complex formation obeys Eq. (3)
and that for such a system the value of k ranges from \( 1.4\times 10^{6}M^{-1}s^{-1} \) to \( 4.2\times 10^{6}M^{-1}s^{-1} \). 

An integrated form of Eq. (3) is given by 

\hfill{} \( (P_{0}-S_{0})^{-1}ln[\{S_{0}(P_{0}-[PS]_{t})\}/\{P_{0}(S_{0}-[PS]_{t})\}]=k(t-t_{0}) \) \hfill{}(4)

which expresses the time elapsed \( (t-t_{0}) \) from the onset of the reaction
at \( t=t_{0} \) in terms of the initial concentrations \( S_{0} \), \( P_{0} \) and the instantaneous
concentration \( [PS]_{t} \). 

We shall now indicate the way Eq.(4) can be used for designing an
\em optimal \em experimental arrangement that would permit us to estimate
with controllable accuracy the time at which the attachment of photolyase
to the uv-damaged DNA molecules begins (the instant \( t=t_{0} \) of Eq.(4)).
Typically, a certain fraction of the uv-exposed DNA molecules will
be damaged by the actual absorption of uv photons. Among the photolyase
molecules moving about randomly in the aqueous solution, the ones
which are sufficiently close to the uv-damaged DNA molecules get attached
to the specific sites. Then \( [PS]_{t} \) increases gradually with time, reaching
a plateau after a certain time following Eq.(4). For our purpose,
the relevant parameters \( P_{0} \), \( S_{0} \) need to be chosen such that \em sufficiently
long period \em elapses before the plateau region sets in. This will
enable withdrawing a number of aliquots from the sample and determining
the fraction of photolyase attached to uv-damaged DNA sites at various
time points. 

The absorption of uv photons by DNA follows Beer's law given by 

\hfill{}\( ln\left( \frac{I}{I_{0}}\right) =A=\varepsilon c_{m}L \) \hfill{}(5) 

where \( I_{0} \) is the incident intensity, I is the intensity emerging from
the sample, A is known as the absorbance of the sample, \( \varepsilon  \) is the extinction
coefficient, \( c_{m} \) is the molar concentration and L is the pathlength
traversed by the photons. For a given sample, both A and \( \varepsilon  \) depend
on the wavelength of incident radiation and nature of the solvent.

For the sake of concreteness, let us choose a \( 10^{-10}M \) concentration for
the aqueous soluti on of synthetic DNA of 10 base pairs (1 base pair
\( \cong  \)\( 600 \)\( m_{H} \) where \( m_{H} \) is the mass of a hydrogen atom) containing a single potential
pyrimidine dimer formation site (two adjacent thymines). For such
a solution, \( \varepsilon \cong 10^{5}M^{-1}cm^{-1} \)corresponding to the uv wavelength \( \sim  \)250-270 nm; for a
pathlength of L = 10 cm, A turns out to be \( 10^{-4} \). Note that higher concentrations
of DNA would make the attachment of photolyase too fast for our purpose.

It is operationally convenient to start with the DNA site concentration
far in excess of the photolyase, i.e., \( S_{0}\gg P_{0},[PS]_{t} \)whence the second order rate
law effectively simplifies to a pseudo first order one. This implies
that in Eq.(3) the product of k and \( S_{0} \)(\( \cong [S]_{t}) \)can be taken as the new first
order rate constant. With a \( 10^{-10}M \) concentration for the 10 base pair DNA,
one may choose a far less initial concentration of photolyase, say,
\( 10^{-12}M \) whose changes due to binding with DNA can be determined with reasonable
accuracy by using carbon radioisotope labelled photolyase [23]. 

We shall now estimate the number of uv photons required to completely
convert all the adjacent thymines into pyrimidine dimers (potential
photolyase attachment sites) in the 10 base pair DNA molecules. We
consider that these molecules are in an aqueous solution contained
within a size of 1mm x 1mm x 10 cm (pathlength). Since \( A=10^{-4} \), this means
that 0.023\% of the incident uv photons are absorbed by such a solution.
However, absorption of an uv photon does \em not \em necessarily lead
to the formation of a pyrimidine dimer in a DNA molecule. Like any
photoreaction, the formation of pyrimidine dimers depends on a quantum
yield \( \phi  \) which is the ratio of the number of photons utilised in forming
pyrimidine dimers to the number of photons actually absorbed. For
a quantum yield \( \phi =0.015 \) observed in the case of polynucleotides [24], it
is calculated that \( 1.74\times 10^{15} \) incident uv photons are required to convert \em all
\em the adjacent thymines into pyrimidine dimers in the solution whose
concentration and size are indicated above. This means that if the
source emits a pulse of \( 10^{9}\gamma  \) photons, the resulting \( 10^{15} \) uv photons are
sufficient to convert a sizeable number of DNA molecules into dimers
or potential photolyase attachment sites. 

Now let us suppose that the above solution containing \( 10^{-10}M \) of 10 base
pair DNA and \( 10^{-12}M \) of photolyase is left exposed (in conjunction with
the CsI arrangement mentioned earlier) for a certain time to a source
which has a probability of emitting a pulse of \( 10^{9}\gamma  \) photons within that
time interval. Then to ``read'' the relevant outcome and to know ``when''
the outcome was registered, the above solution needs to be subjected
to the following procedure. By withdrawing aliquots (random portions
of the sample) from the solution at various times and running them
through a polyacrylamide gel [25], two spatially separated radioactively
labelled bands will result \em if \em the source has emitted photons
: one band corresponding to the \em unbound \em photolyase \( (P_{0}-[PS]_{t}) \) and the
other to the photolyase \em bound \em to uv-damaged DNA \( ([PS]_{t}) \) which is
\em heavier\em . The proportion of photolyase in these bands within
polyacrylamide gel can be determined by measuring the radioactive
counts from them. Then as indicated below, \( t_{0} \) can be estimated by using
Eq.(4) on the basis of such measurements for any pair of instants
\( t_{1} \), \( t_{2} \) provided \( S_{0} \) is much greater than \( P_{0} \) (pseudo first order reaction).
This condition is satisfied by the concentrations chosen so that Eq.(4)
simplifies to 

\hfill{}\( ln[(P_{0}-[PS]_{t})/P_{0}]=-S_{0}k(t-t_{0}) \)\hfill{}(6)

Using Eq.(6) for any pair of instants \( t_{1} \), \( t_{2} \) we get the following relation

\hfill{}\( ln[(P_{0}-[PS]_{t_{1}})/(P_{0}-[PS]_{t_{2}})]=S_{0}k(t_{2}-t_{1}) \)\hfill{}(7)

Since left hand side of Eq.(7) is experimentally determined by the
radioactive counts from photolyase bands within the polyacrylamide
gel, (\( S_{0}k \)) can thus be calculated. This would enable one to estimate
\( t_{0} \) from Eq.(6) for either \( t_{1} \)or \( t_{2} \). The half life associated with the
pseudo first order process corresponding to Eq.(6) is given by \( ln2/(S_{0}k) \).
Substituting the values for \( S_{0} \) and k mentioned earlier, this half life
ranges from 57 to 83 minutes. Thus it is feasible to ensure a reasonably
sufficient time before \( [PS]_{t} \) reaches a plateau. One can therefore appropriately
\em increase \em the accuracy in estimating \( t_{0} \) by having more withdrawals
and data at different times over the region where \( [PS]_{t} \) varies with time
significantly. 

It should be noted that the operational significance of \( t \), \( t_{1} \), \( t_{2} \) occurring
in Eqs. (4), (6), (7) is that they refer to the instants of \em applying
\em the withdrawn aliquots to the gel (\em not \em the instants of
their withdrawal from the solution). This is because \( t \), \( t_{1} \),\( t_{2} \) are the
instants \em till \em which the photolyase attachment goes on before
the separation between unbound and bound photolyase is made within
the gel. 

In the situation where the condition for the pseudo first order rate
law is not satisfied, one can still use Eq.(4) for computing \( t_{0} \) but
then the procedure is more complicated requiring data at three different
instants. We also note that the time required for the formation of
uv induced pyrimidine dimer in a DNA molecule is known to be exceedingly
small, \( \sim 10^{-14}s \) [26]; hence this period can be ignored in interpreting that
\( t_{0} \) signifies the instant from which the relevant outcome is registered.

To sum up, our preceding arguments show that the phenomenologically
justified Rate Equation (whose various forms are Eqs. (3), (4), and
(6)) entails \em empirically \em \em determinable \( t_{0} \). \em In the context
of the quantum measurement problem, \( t_{0} \) signifies the instant from which
the outcome that a source has emitted photons \em is knowable \em in
the sense that at \em any \em \( t\geq t_{0} \), suitable observations can be made
to ``read'' this outcome. In other words, \em \( t_{0} \) \em signifies the \em onset
\em of a \em discernible heterogeneity \em (mixed state) from an initial
homogeneity (pure state). Thus \( t_{0} \) has an objective factual significance
which needs to be accommodated in a consistent way within the framework
of quantum mechanics. To what extent the various models / interpretations
of quantum mechanics can achieve this for such examples of quantum
measurement calls for careful scrutiny. In particular, if a wave function
is taken to provide a ``complete'' description of the state of a system,
it becomes necessary to explain in detail the dynamical process by
which the superposition of states given by Eq.(2) starts \em reducing
\em to an actual mixed state of two terms (\( \psi _{e}\phi _{A} \) and \( \psi _{0}\phi _{0} \)) in Eq. (2), \em concomitant
\em with the onset of photolyase attachment to uv-absorbed DNA molecules.

Our scheme may be considered \em analogous \em to retrodictively inferring
from suitable observations the \em instant \em at which a cat has
expired upon the emission of photons in Schroedinger's famous thought
experiment [27], popularly known as the ``cat paradox''. Our suggested
new twist to the ``cat paradox'' makes it amenable to \em controlled
\em experimental studies using biomolecular systems (in our example,
photolyase attached DNA is analogous to the ``dead'' state of a cat
in the context of ``cat paradox''). Investigations along this direction
have the potentiality to provide useful empirical clues and fresh
insights into the quantum measurement problem, particularly because
of \em mesoscopic \em sizes and masses of the biomolecules involved
[28]. 

DH is grateful to Roger Penrose for raising insightful questions concerning
our earlier work [Ref. 21] which helped in the formulation of the
present work. DH also thanks John Corbett for stimulating discussions.
RC acknowledges research grant by BRNS, Govt. of India (4/5/95-R \&
D-II/703). 

\( \smallskip  \)

\bfseries REFERENCES AND NOTES \mdseries 

\( \smallskip  \)

1. A.J. Leggett, in \em Quantum Implications\em , B.J. Hiley, \& D.
Peat, Eds.(Routledge \& Kegan Paul, London, 1987), pp. 85-104. 

2. J.S. Bell, \em Speakable and Unspeakable in Quantum Mechanics \em (Cambridge
University Press, Cambridge, 1987).

3. B. D'Espagnat, \em Veiled Reality - An Analysis of Present -Day
Quantum Mechanical Concepts \em (Addison - Wesley, Reading, 1994).

4. J.T. Cushing, \em Quantum Mechanics - Historical Contingency and
the Copenhagen Hegemony \em (University of Chicago Press, Chicago,
1994). 

5. A. Whitaker, \em Einstein, Bohr and the Quantum Dilemma \em (Cambridge
University Press, Cambridge, 1996). 

6. M. Namiki, S. Pascazio, and H. Nakazato, \em Decoherence and Quantum
\em \em Measurement \em (World Scientific, Singapore, 1997). 

7. D. Home, \em Conceptual Foundations of Quantum Physics - An Overview
from Modern Perspectives \em (Plenum Press, New York, 1997). 

8. P. Mittelstaedt, \em The Interpretation of Quantum Mechanics and
the Measurement Process \em (Cambridge University Press, Cambridge
1998). 

9. S. Weinberg, \em Dreams of a Final Theory\em , (Vintage, London,
1993), p. 64.

10. Coupling of measuring apparatus with environmental degrees of
freedom is often invoked to argue that coherence effects embodied
in the pure state entangled form Eq.(1) are difficult to observe in
practice (\em effective \em \em decoherence\em ). The underlying contention
is that the pure state (1) behaves \em as if \em it were a mixed state.
Whether such ``effective \em \em decoherence'' between different outcomes
is \em sufficient \em to ensure their \em actual \em distinguishability
is highly debatable [1-4,7]. 

11. P. Holland, \em The Quantum Theory of Motion \em (Cambridge University
Press, Cambridge, 1993). 

12. D. Bohm, and B.J. Hiley, \em The Undivided Universe \em (Routledge,
London, 1993).

13. R. Griffiths, \em J. Stat. Phys\em . \bfseries 36\mdseries ,
219 (1984). 

14. R. Omnes, \em The Interpretation of Quantum Mechanics \em (Princeton
University Press, Princeton, 1994); \em Understanding Quantum Mechanics
\em (Princeton University Press, Princeton, 1999). 

15. N. Gisin, in \em Fundamental Problems in Quantum Theory\em , D.M.
Greenberger, and A. Zeilinger, Eds. (Annals of New York Academy of
Sciences, New York, 1995), pp. 524-533. 

16. G.C. Ghirardi, A. Rimini, and T. Weber, \em Phys. Rev. D \em \bfseries 34\mdseries ,
470 (1986); P. Pearle, \em Phys. Rev. A \bfseries \em 39\mdseries ,
2277 (1989). 

17. G.C. Ghirardi, R. Grassi and A. Rimini, \em Phys. Rev. A \em \bfseries 42\mdseries ,
1057 (1990); L. Diosi, \em Phys. Rev. A \em \bfseries 40\mdseries ,
1165 (1989). 

18. G.C. Ghirardi, in \em Fundamental Problems in Quantum Theory\em ,
D.M. Greenberger, and A. Zeilinger. Eds. (Annals of New York Academy
of Sciences, New York, 1995), pp. 506-523; G.C. Ghirardi, and R. Grassi,
in \em Bohmian Mechanics and \em \em Quantum Theory : An Appraisal\em ,
J.T. Cushing, A. Fine, and S. Goldstein, Eds. (Kluwer, Dordrecht,
1996), pp. 353-377. 

19. R. Penrose, \em Gen. Rel. and Gravitation\em , \bfseries 28\mdseries ,
581 (1996). 

20. A. Shimony, in \em Philosophical Consequences of Quantum Theory,
\em J.T. Cushing and E. McMullin, eds. (University of Notre Dame Press
, Notre Dame , 1989), p. 61. 

21. D. Home and R. Chattopadhyaya, \em Phys. Rev. Lett\em ., \bfseries 76\mdseries ,
2836 (1996). 

22. P. Pearle and E. Squires, \em Phys. Rev. Lett\em ., \bfseries 80\mdseries ,
1348 (1998); see also Reply by D. Home and R. Chattopadhyaya, \em Phys.
Rev. Lett\em ., \bfseries 80\mdseries , 1348 (1998). 

23. G.B. Sancar et al., \em J. Biol. Chem\em ., \bfseries 262\mdseries ,
478 (1987). 

24. M.H. Patrick and R.D. Rahn, in \em Photochemistry and Photobiology
of Nucleic Acids, \em S.Y. Wang, ed., (Academic Press, New York, 1976),
Vol II, pp. 35-145. 

25. I. Husain and A. Sancar, \em Nucl. Acids Res\em ., \bfseries 15\mdseries ,
1109 (1987). 

26. R.P. Wayne, in \em Principles and Applications of Photochemistry\em ,
(Oxford University Press, New York, 1998), pp. 30-32. 

27. E. Schroedinger, \em Naturwissenschaften\em , \bfseries 23\mdseries ,
804 (1935). 

28. For a quantitative idea about sizes and masses of the relevant
biomolecules, see H.W. Park, S.T. Kim, A. Sancar and J. Deisenhofer,
\em Science\em , \bfseries 268\mdseries , 1866 (1995). 

\end{document}